\documentclass[aps,pra,twocolumn,preprintnumbers]{revtex4}
\usepackage{amsmath,amssymb,bm,graphicx,subfigure}

\begin{document}

\title{Effective single-band models for strongly interacting fermions in an
optical lattice}
\author{J. P. Kestner\footnote{\textit{Current address: Condensed Matter Theory Center, Department of Physics, University of Maryland, College Park, MD 20742.}}, L.-M. Duan}
\affiliation{Department of Physics and MCTP, University of Michigan, Ann Arbor, MI 48109}

\begin{abstract}
To test effective Hamiltonians for strongly interacting fermions in an optical lattice, we numerically find the energy
spectrum for two fermions interacting across a Feshbach resonance in a double well potential. From the spectrum, we
determine the range of detunings for which the system can be described by an effective lattice model, and how the model
parameters are related to the experimental parameters. We find that for a range of strong interactions the system is
well described by an effective $t-J$ model, and the effective superexchange term, $J$, can be smoothly tuned through
zero on either side of unitarity. Right at and around unitarity, an effective one-band general Hubbard model is
appropriate, with a finite and small on-site energy, due to a lattice-induced anharmonic coupling between atoms at the
scattering threshold and a weakly bound Feshbach molecule in an excited center of mass state.
\end{abstract}

\maketitle

\section{Introduction}

Systems of ultracold fermionic atoms interacting in an optical lattice potential via a Feshbach resonance provide
remarkable opportunities to realize a zoo of lattice Hamiltonians in a clean and controllable fashion \cite{Jaksch05,
Duan05, Lewenstein07}. This allows a new tool to study many lattice models familiar from condensed matter physics such
as the Hubbard, $t-J$, and $XXZ$ models, to name a few. In addition to the prospect of studying these paradigmatic
models experimentally in the absence of unwanted complications, ultracold gases provide the exciting possibility of
gaining new insight into the proper description of strongly correlated systems.

While it is well-known that a weakly interacting gas in an optical lattice can be described by the one-band Hubbard
model \cite{H02, Jaksch05}, the situation for strongly interacting gas near a Feshbach resonance is much more
complicated. In the strongly interacting regime, the conventional assumptions for derivation of the one-band Hubbard
model for two-component fermions obviously no longer applies, since the on-site interaction energy becomes greater than
the bandgap of the lattice and the off-site interaction gets comparable with the atomic tunneling rate \cite{Duan05,
Ho06}. However, Refs. \cite{Duan05, Duan08} provide general arguments to show that in this case we can still derive an
effective single-band lattice Hamiltonian, in the form of either a general Hubbard model with possibly particle
correlated hopping rates, or a $t-J$ or $XXZ$ model under different limiting situations. This simplification relies on
several observations: first, with the assumption that the average number of atoms per lattice site $\bar{n} \leq 2$, it
is unlikely for more than two atoms to occupy the same site because they are energetically unfavorable due to the Pauli
exclusion and the strong on-site interaction \cite{Kestner07}. Second, with two atoms on the same site, although they
will populate many lattice bands due to strong interaction, the two-atom eigen-levels (called the dressed molecules)
have large energy splitting between them \cite{Busch,05, Duan05, Ho06, Kestner06, Werner05}, and for low-temperature
physics, only one of these dressed molecule levels will be relevant (the other levels can be adiabatically eliminated in the derivation). So the effective Hilbert space is severely restricted: each site may be empty, or populated with either one atom, or one dressed molecule with a fixed internal state. Based on restriction of the effective Hilbert space and
the SU(2) symmetry for the underlying physical process, it is derived in Ref. \cite{Duan08} that the effective lattice
Hamiltonian takes the form of a general Hubbard model with possibly particle correlated hopping rates. The multiple-band population and the off-site interaction are taken into account through renormalization of the effective parameters in
the final single-band model. In different interaction regions for the atoms (controlled by the external magnetic field
through the Feshbach resonance), the dimension of the effective Hilbert space on each lattice site can be further reduced, leading to an effective $t-J$ model for the atoms or an $XXZ$ model for the dressed molecules \cite{Duan05, Duan08}. Note though, that in these regions the $t-J$ or the $XXZ$ models in general do not arise from the Hubbard (or the general Hubbard) model via perturbation theory, and they should be taken as the basic models for the underlying physical process.

This paper serves two purposes: First, we provide numerical tests for the effective lattice Hamiltonians applied to
the two-body problem. The effective Hamiltonian should describe two-body physics as well as many-body physics. For the
two-body problem, we can solve it exactly in a double well lattice from the basic strongly interacting field Hamiltonian and calculate the low-energy eigen-levels of the system. We then compare the results from the exact numerical solutions
with prediction from the effective single-band models. This comparison determines the validity region for each effective single-band model. We note that the two-body strongly interacting problem has also been solved for a three-dimensional lattice in the BCS region \cite{Zoller}, for a one-dimensional
lattice \cite{Stringari}, and for two bosons in a triple-well potential \cite{Schneider09}, based on different kinds of
numerical methods. Second, with comparison of the two-body physics, we can microscopically derive the parameters for the effective single-band models from the relevant experimental parameters. The determination of the effective model
parameters is very important for application of these models to do quantitative calculations to compare theory and
experiments. We find that for a wide range of strong interaction, the system is well described by an effective $t-J$
model, and the effective superexchange term, $J$, can be smoothly tuned through zero on either side of unitarity. Right
at and around unitarity, an effective one-band general Hubbard model is appropriate, with a finite and small effective
on-site energy, due to a lattice-induced anharmonic coupling between atoms at the scattering threshold and a weakly
bound Feshbach molecule in an excited center of mass state (the so-called anharmonic induced resonances, see Ref.
\cite{Kestner09}).

This paper is arranged as follows: In Sec. II we specify our numerical approach for exact solution of the two-body
problem in a double-well potential, which is based on the stochastic variational method \cite{Suzuki}, a widely-used
method in nuclear physics for solution of the few-body problems. The main results of this paper are shown in Sec. III.
First, we specify the validity regions for each effective single-band lattice model across the whole interaction region, from the BEC limit, to resonance, and to the BCS limit. Second, we derive the renormalized parameters for these
effective lattice models. The values for some of these parameters are pretty counter-intuitive, and in such cases we
discuss their physical origin and consequence.

\section{Methods}

\begin{figure}[tbp]
\subfigure  {\includegraphics[height=.35\columnwidth]
        {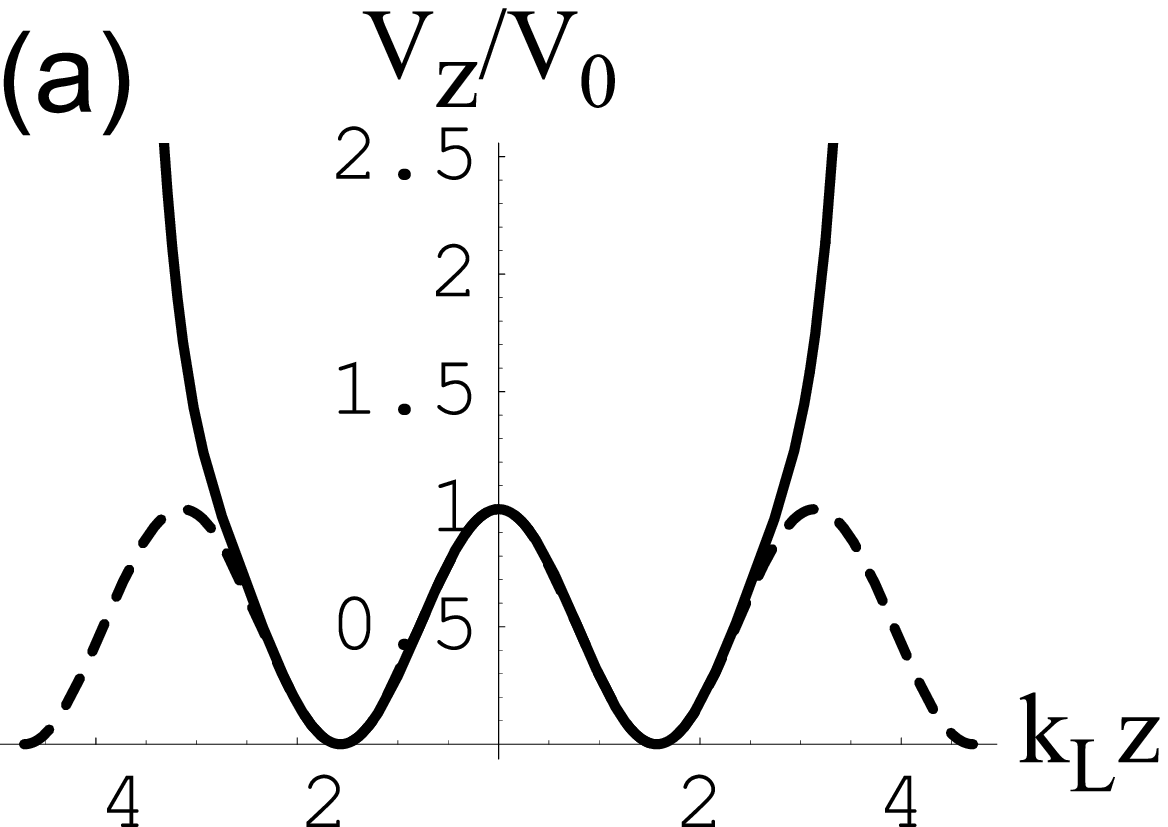}}
\subfigure  {\includegraphics[height=.35\columnwidth]
        {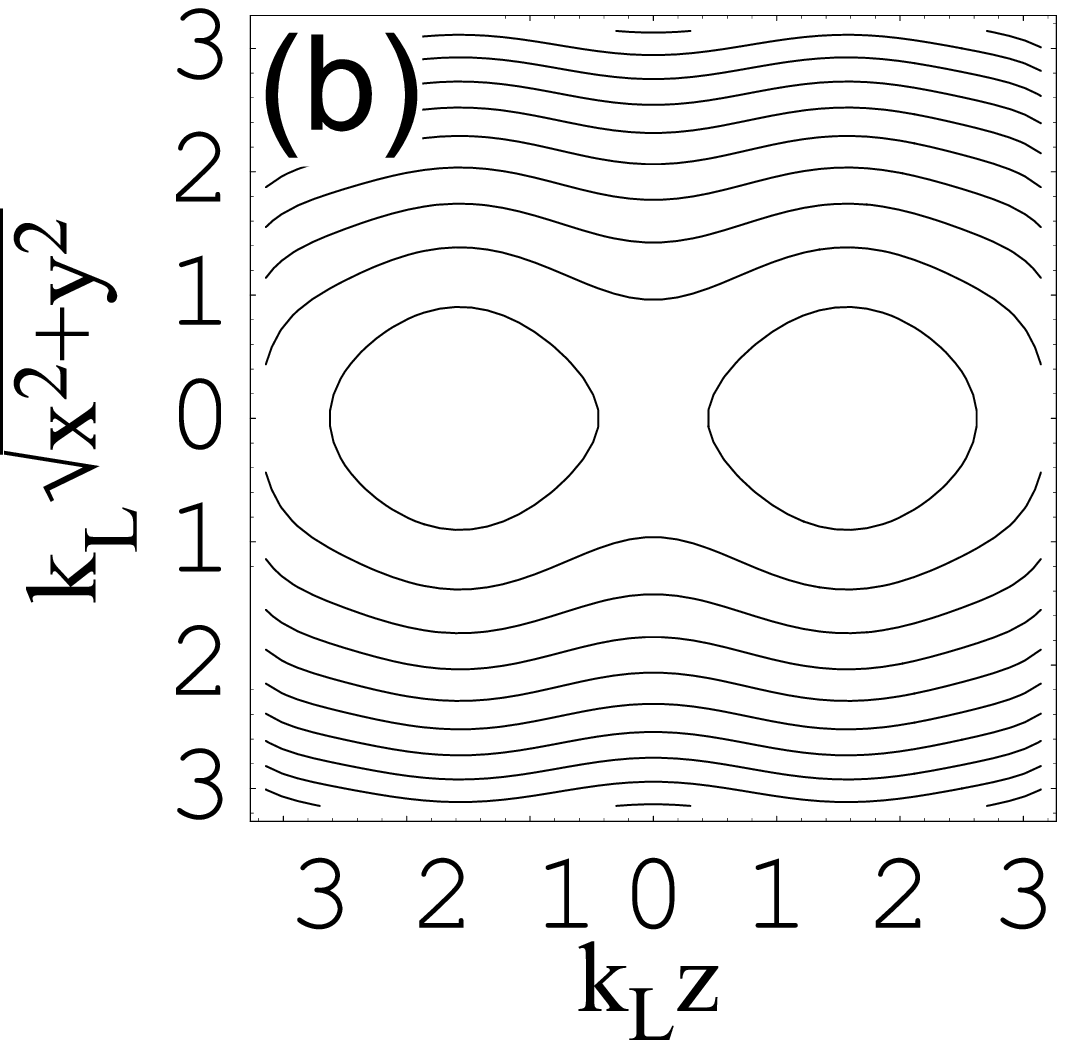}}
\caption{(a) The double-well potential along $z$ modeling a periodic potential; (b) contour plot of the locally
isotropic 3D double-well potential.} \label{fig:dblwell}
\end{figure}

We consider two distinguishable fermions of mass $m$ in an external potential $V\left( \mathbf{x}\right) $, interacting
via a short range potential $U\left( r\right) $ characterized by its s-wave scattering length, $a_{s}$. In experiments,
the external potential is typically the sum of a periodic potential $V_{0}\prod_{i=1}^{3}\cos ^{2}k_{L}x_{i}$ and a
harmonic confining potential. To make our calculations easier we will model this by considering a double-well potential
along the $z$-axis (formed by Taylor expanding $\cos ^{2}k_{L}z$) and a harmonic potential in the other two transverse
directions, with the frequency, $\omega $, chosen such that the potential is locally isotropic at the bottom of each
well, as shown in Fig.~\ref{fig:dblwell}. In fact, this double-well problem is interesting in its own right, as it is
relevant to experiments with gases confined in optical superlattices \cite{dblwell}. Since the barrier is the same as
for the full lattice, the hopping rates should be nearly unaffected. Also, the on-site interaction energy should not be
affected qualitatively. The only relevant information that the double-well approximation renders inaccessible is rates
for any next-nearest neighbor process. The effective single-band lattice Hamiltonians we are comparing belong to the
tight-binding models with interaction only among the nearest neighbors, so it is enough to consider a double-well
potential to test these Hamiltonians and their parameters.

Due to the harmonicity of the transverse trap, the center of mass (c.m.)\ motion in the transverse direction separates
out and is thus neglected in the rest of the discussion. However, along the axis of the double-well, the c.m.\ motion is not separable from the relative motion. The two atom system then has three relevant coordinates: the relative
coordinates $z$ and $\rho$, along the axial and transverse directions, respectively, and $Z$, the axial c.m.\
coordinate. In terms of these coordinates, the external potential approximating an optical lattice of depth $V_0$ is $V
\left( \rho, z, Z \right) = V \left( \rho \right) + V \left( z, Z \right)$, where
\begin{multline}\label{eq:V}
V\! \left( \rho \right) = V_0 k_L^2 \rho^2/2
\\
V\! \left( z, Z \right) = V_0 \sum_{ \substack{ n=0 \\ \pm} }^6 \frac{ \left( -4 \right)^n \Gamma \left( 1 - 2n \right) }{ \Gamma \left( 1- 4n \right) \Gamma \left( 1 + 4n \right) } k_L^{2n} \left( Z \pm \frac{z}{2} \right)^{2n}
\end{multline}
and $\Gamma \left(x\right)$ is the Euler gamma function.

The exact form of the interaction, $U\left( \sqrt{\rho ^{2}+z^{2}}\right) $, is irrelevant in the low-energy limit as
long as its range is much smaller than the average interatomic distance and the trap length scale, and most analytical
treatments use a zero-range pseudopotential. Numerically, it is easiest to use a finite-range attractive Gaussian
interaction $U\left( r\right) =-U_{0}\exp \left( -r^{2}/r_{0}^{2}\right) $, where we typically take
$r_{0}=0.05\sqrt{\hbar /m\omega }$. Finite-range effects should be negligible for such small values of $r_{0}$, and we
have verified this by repeating the calculations with $r_{0}=0.1\sqrt{\hbar /m\omega }$. The free space scattering
length is varied by adjusting the strength of the interaction, $U_{0}$.

Adopting units such that $k_L = 1$ and $E_R = \hbar^2 k_L^2/2m = 1$, the Hamiltonian may be written as
\begin{multline}  \label{eq:Hr}
H=-\frac{2}{\rho} \frac{\partial}{\partial \rho} \rho \frac{\partial}{\partial \rho} -2\frac{\partial^2}{\partial z^2} - \frac{1}{2}\frac{\partial}{\partial Z^2}
\\
+2\frac{m^2_{\ell}}{\rho^2}+ V\left( \rho ,z,Z\right) -U_{0}e^{-\left( z^{2}+\rho ^{2}\right)
/r_{0}^{2}}
\end{multline}
where $m_{\ell}$ is the relative angular momentum, which is a good quantum number due to axial symmetry.  In the following we will only consider $m_{\ell}=0$, since, in the limit as $r_0$ goes to zero, the interaction does not affect states with $m_{\ell} \neq 0$.

We find the low-lying states of the system using a stochastic variational method \cite{Suzuki} recently introduced to
the ultracold gas community \cite{Stecher07}. In this approach, the variational wavefunction takes the form
\begin{equation}  \label{eq:psi}
\Psi \left( \rho, z, Z \right) = \sum_i^N \alpha_i \exp \left( -\rho^2/a_i^2 - z^2/b_i^2 - Z^2/c_i^2 \right) ,
\end{equation}
where $\alpha$ is a linear variational parameter, $\{ a, b, c \}$ are nonlinear variational parameters which define the
basis elements, and $N$ is the size of the basis set. The nonlinear parameters are selected from stochastically
generated pools of candidates to minimize the variational energy $\langle \Psi_{abc} |H| \Psi_{abc} \rangle / \langle
\Psi_{abc} | \Psi_{abc} \rangle$. The basic algorithm is as follows: starting with a set of $N-1$ basis states,

\begin{description}
\item[1)] a pool of (in our calculations) $25$ new basis states is randomly generated, each defined by a given value
    of $\{a_{i},b_{i},c_{i}\}$;

\item[2)] for each of the $25$ possible $N$-dimensional basis sets formed by adding one basis from the candidate
    pool, the energy is minimized with respect to $\alpha $;

\item[3)] the new basis set that yields the lowest energy is kept and the previous steps are repeated until the
    basis size, $N$, increases to the desired number.
\end{description}

Once every few iterations, the existing basis set is optimized by the following refining process: starting with a set of
$N$ basis states and $n=1$,

\begin{description}
\item[A)] a pool of $25$ replacement basis states is randomly generated, each defined by a given value of
    $\{a_{n},b_{n},c_{n}\}$;

\item[B)] for each of the $25$ possible $N$-dimensional basis sets formed by replacing the $n^{\text{th}}$ old basis
    state with a new one from the candidate pool, the energy is minimized with respect to $\alpha $;

\item[C)] if the lowest of these $25$ energies is lower than the current variational energy, the $n^{\text{th}}$ old
    basis state is replaced by the new optimal one and the previous steps are repeated for $n=1...N$.
\end{description}

The great advantage of this stochastic variational method is that it is able to avoid getting stuck in the local minima of the energy landscape that plague deterministic variational methods.  Unlike Monte Carlo algorithms, though, there is no well-characterized statistical error.  The resulting variational energies set upper bounds on the true eigenenergies, but we cannot attach a rigorous error estimate; we only check that the results seem well-converged.  For a detailed discussion of the method, see Ref.~\cite{Suzuki}.

We typically achieved fairly good convergence for $N\sim 300$. Although in principle the nonlinear basis optimization
must be performed for each value of $a_{s}$, actually the basis set does not change too much as one sweeps across
resonance except to include narrower and narrower Gaussians for positive $a_{s}$ where deeply bound molecules form.
Apart from deeply bound states, the change in the wavefunction is mainly due to changing the expansion coefficients,
$\alpha $. To save computational time then, we performed the nonlinear basis optimization for four different values of
$ a_{s}$ across resonance, joined the four optimized basis sets, and simply minimized the energy with respect to
$\alpha $ using the resultant basis set of about $1200$ elements for all values of $a_{s}$. As a result, very deeply
bound energy levels may not be fully converged, but we are only interested in the energy range around the lowest
non-interacting levels. In this range, our results appear to be converged.

\section{Results}

\subsection{Regions of model validity}

\begin{figure}[tbp]
\subfigure  {\includegraphics[width=1\columnwidth]
        {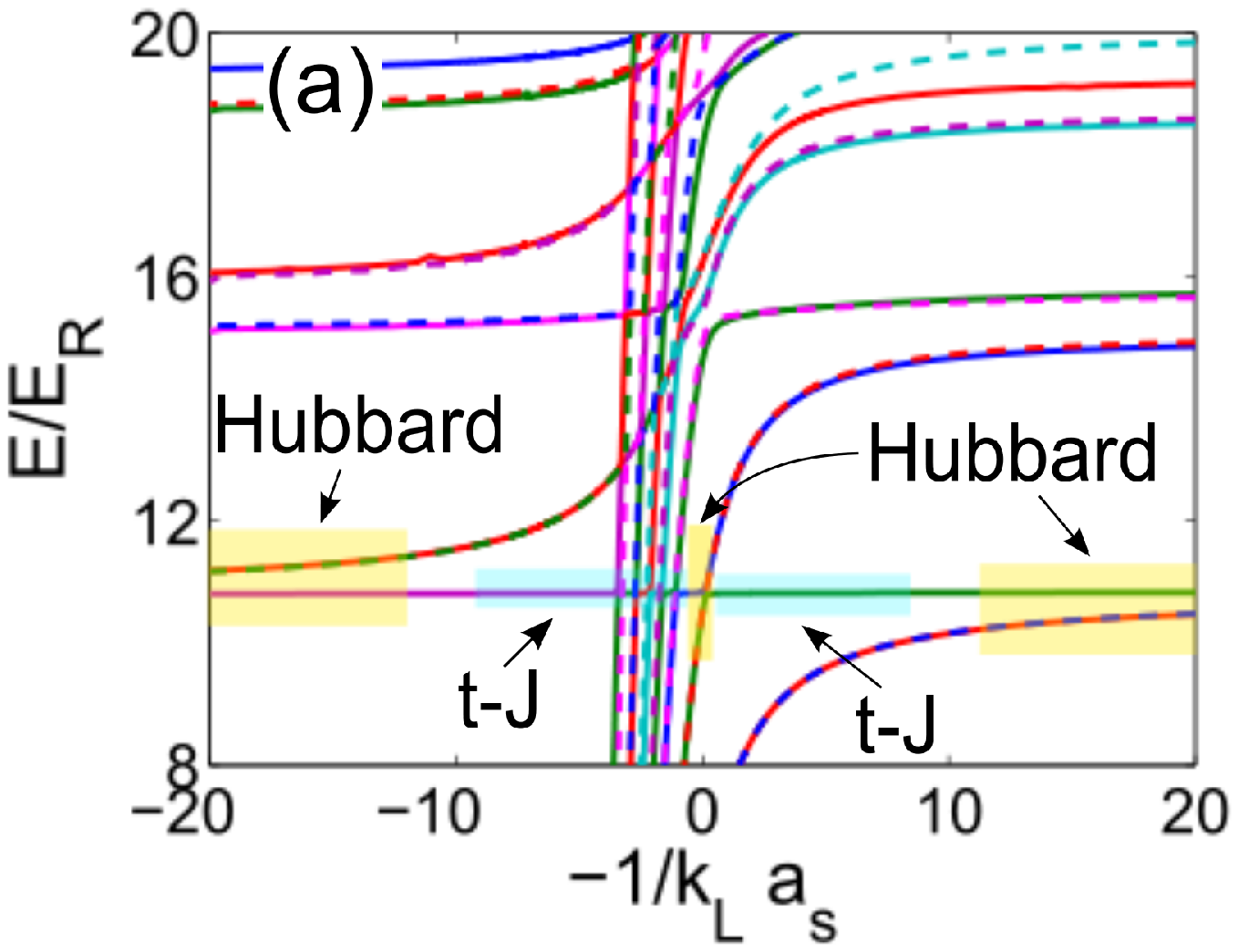}\label{fig:spectrum4a}}
\subfigure  {\includegraphics[width=1\columnwidth]
        {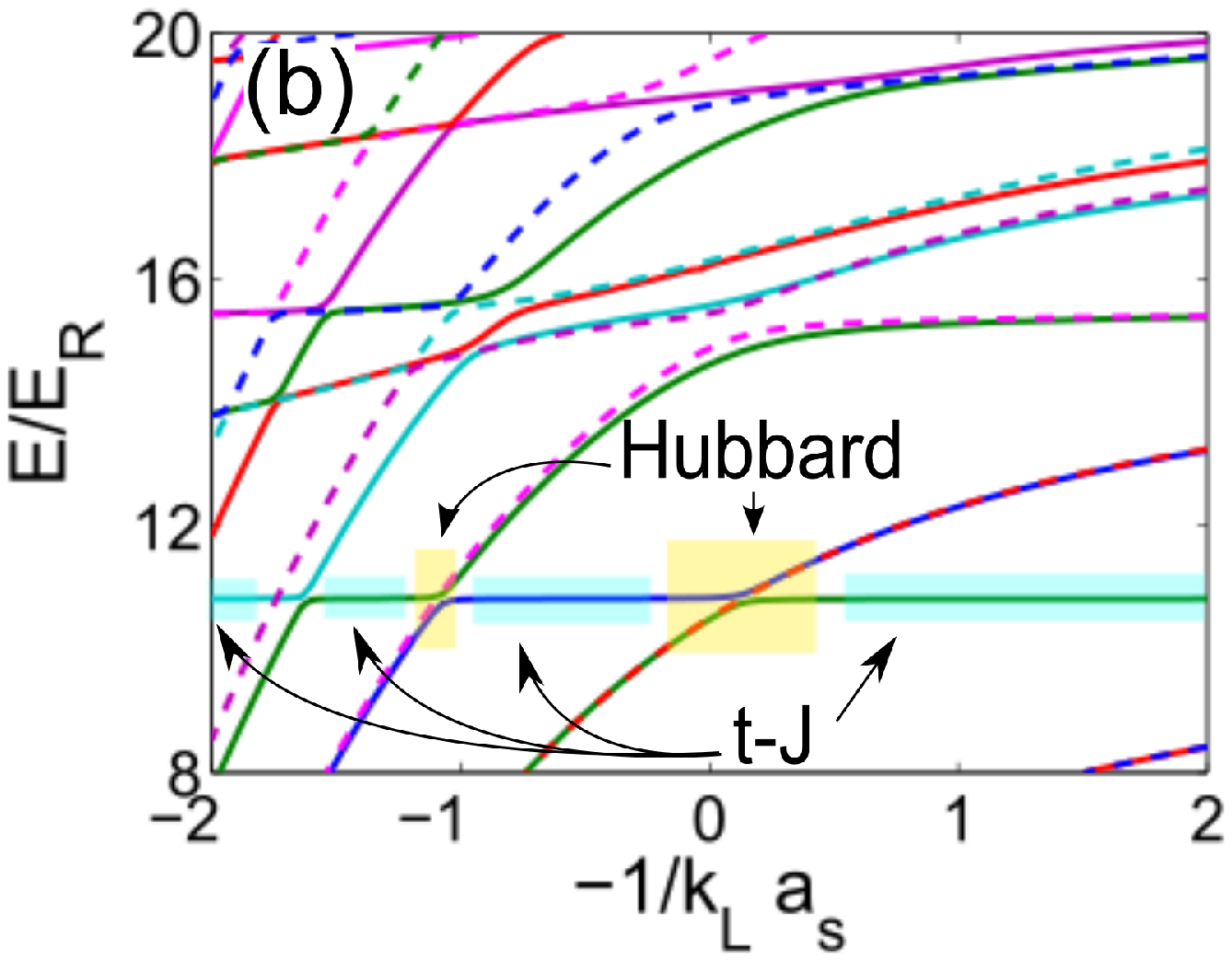}\label{fig:spectrum4b}}
\caption{(Color online.) (a) Spectrum of two interacting atoms in a three-dimensional double-well potential vs. inverse
free space scattering length with $V_{0}=8E_{R}$. Solid (dashed) lines correspond to states of even (odd) symmetry in
$Z$. Only the first few plunging levels are shown. (b) Close-up of the strongly interacting region. The noninteracting
states odd in $z$ are not shown.} \label{fig:spectrum4}
\end{figure}

\begin{figure}[tbp]
\subfigure  {\includegraphics[width=1\columnwidth]
        {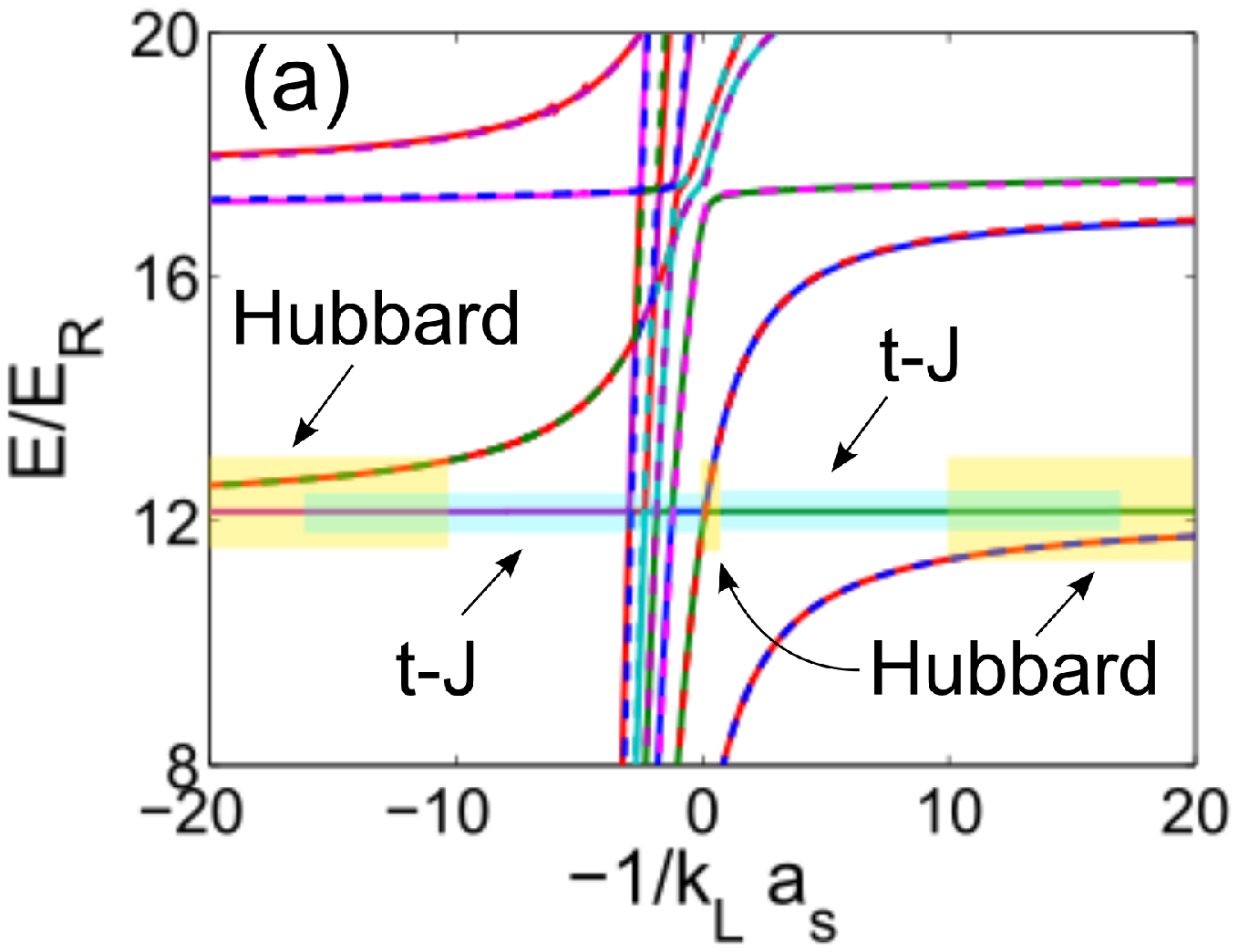}\label{fig:spectrum5a}}
\subfigure  {\includegraphics[width=1\columnwidth]
        {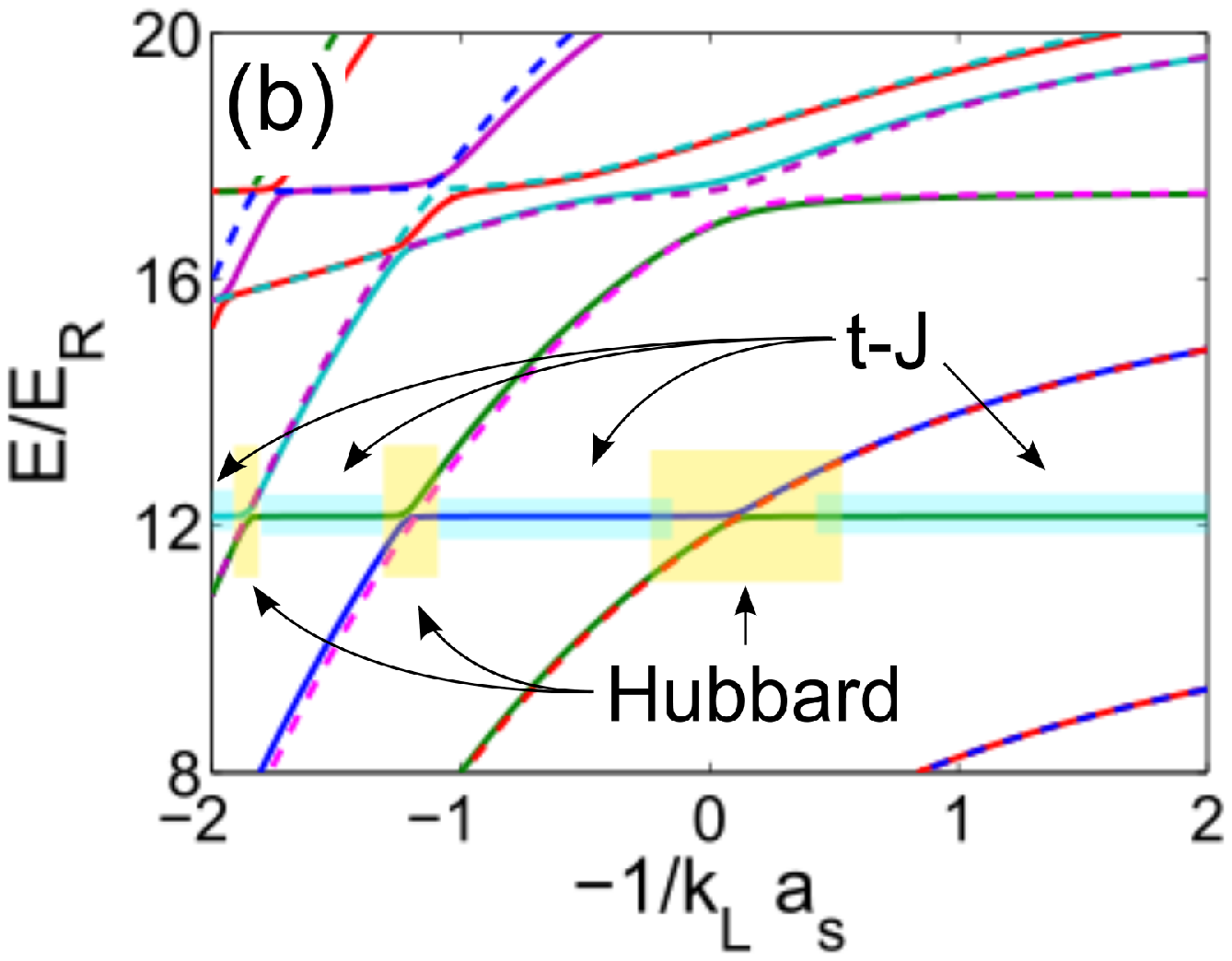}\label{fig:spectrum5b}}
\caption{(Color online.) Same as Fig. \protect\ref{fig:spectrum4} but for $ V_{0}=10E_{R}$.} \label{fig:spectrum5}
\end{figure}

In Figs.~\ref{fig:spectrum4} and \ref{fig:spectrum5}, we show the spectrum of two atoms interacting in a double-well
potential across a Feshbach resonance for $V_{0}=8E_{r}$ and $V_{0}=10E_{r}$. For clarity, we have diabatized the
spectrum for $-1/k_{L}a_{s}<-4$, omitting the plunging levels. We have also omitted the exactly flat, non-interacting
levels. In the absence of trap anharmonicity, there are three kinds of curves present: plunging levels corresponding to
tightly bound molecules in motionally excited states, flat levels corresponding to atoms in separate wells, and
sigmoidal levels corresponding to interacting extended atom pairs. Levels corresponding to states of similar parity
never cross, instead undergoing a rich set of narrow avoided crossings. These are due to Feshbach-type resonances
induced by the anharmonic coupling of the center of mass and relative motion \cite{Kestner09}. Away from unitarity,
these avoided crossings become even narrower and are irrelevant. Some of our calculations of highly excited even
molecule states are apparently not as well converged as the corresponding calculations for the odd states, since the
even states (solid lines) should be the lowest of each plunging doublet as the molecules become tightly bound and act as a single particle. However, this is not important for our purposes.

To approximate the two-atom, double-well physics with a general Hubbard model requires a separation of energy scales
such that there is a manifold of four energy levels well-separated from all the others, corresponding to a doublet of
doubly occupied states and a doublet of singly occupied states. This condition is satisfied far from unitarity on either side. (On the scale of the plots, it may be hard to distinguish the two levels in the doubly occupied doublet, as their
splitting is on the order of the dimer tunneling energy. The levels in the singly occupied doublet are likewise very
close, with splitting on the order of the superexchange energy, but we have only shown the energy of the singlet state
since the triplet state is odd in the relative coordinate and thus independent of the interaction.) We have marked the
low-energy regions of model validity using a rule of thumb that the energy separation between the relevant low-lying
manifold and the nearest level outside the manifold should be at least five times larger than the energy range of the
manifold. Actually, in enforcing this requirement, we have taken into account that the discrete levels turn into bands
of width $ \sim 4 t$ when extending the two-site potential to an infinite one-dimensional lattice.

The small disconnected regions near unitarity where a general Hubbard model is applicable, shown in
Figs.~\ref{fig:spectrum4b} and \ref{fig:spectrum5b}, are qualitatively different from the weakly interacting regions.
Here the on-site dimers correspond to Feshbach molecules in an excited center of mass band. The usual ground band
Feshbach molecules are far-detuned and irrelevant. The coupling to the relevant singly occupied states (which are in the lowest center of mass band) is facilitated by the anharmonicity of the potential. This is a very interesting phenomenon
whereby the presence of a narrow anharmonicity induced resonance \cite{Kestner09} near the wide free space Feshbach
resonance allows for an effective single-band description where the on-site energy becomes very small instead of
arbitrarily large. As higher excited molecular bands become relevant, eventually the molecule bandwidth becomes
comparable to the bandgap and one can no longer apply a general Hubbard model. This is the case in
Fig.~\ref{fig:spectrum4b} at $ -1/k_{L}a_{s}\sim -1.7$.

Likewise, approximation by a $t-J$ model requires that two singly occupied states are well-separated from the others,
and these regions are also marked in Figs.~\ref{fig:spectrum4} and \ref{fig:spectrum5}. (Again, one of these states is
not shown and would be indistinguishable anyway on the scale of the plot.) Note that the form of the $t-J$ model is
valid at detunings where the general Hubbard model is not. There the $t-J$ model does not come from the usual
perturbative treatment of the Hubbard model for $|U|\gg t$, which gives $J=2t^{2}/U$ \cite{Auerbach}. Instead it is the
fundamental model in that region and there are no \textit{a priori} constraints on the parameters. The regions of
validity shown in Figs.~\ref{fig:spectrum4} and \ref{fig:spectrum5} for the two types of lattice models are one of the
main results of this paper.

We have performed the calculations for lattice depths of $V_{0}=2-10E_{R}$. For $V_{0}\leq 4E_{R}$, a lattice model is
not valid for any value of the scattering length. At $V_{0}=6E_{R}$, the bandgap has increased enough that the lattice
models become valid in regions, similar to those plotted in Fig.~\ref{fig:spectrum4}, but narrower. These regions
quickly widen when the lattice depth is increased to $V_{0}=8E_{R}$. As the lattice is deepened to $V_{0}=10E_{R}$,
the $t-J$ model continues to quickly expand its validity further into the weakly interacting regime, overlapping with
the regions where a Hubbard model is valid. It expands only slightly closer to unitarity. A general Hubbard model also
becomes valid over only a slightly larger region.

\subsection{Model parameters}
\begin{figure}[tbp]
\subfigure  {\includegraphics[width=1\columnwidth]
        {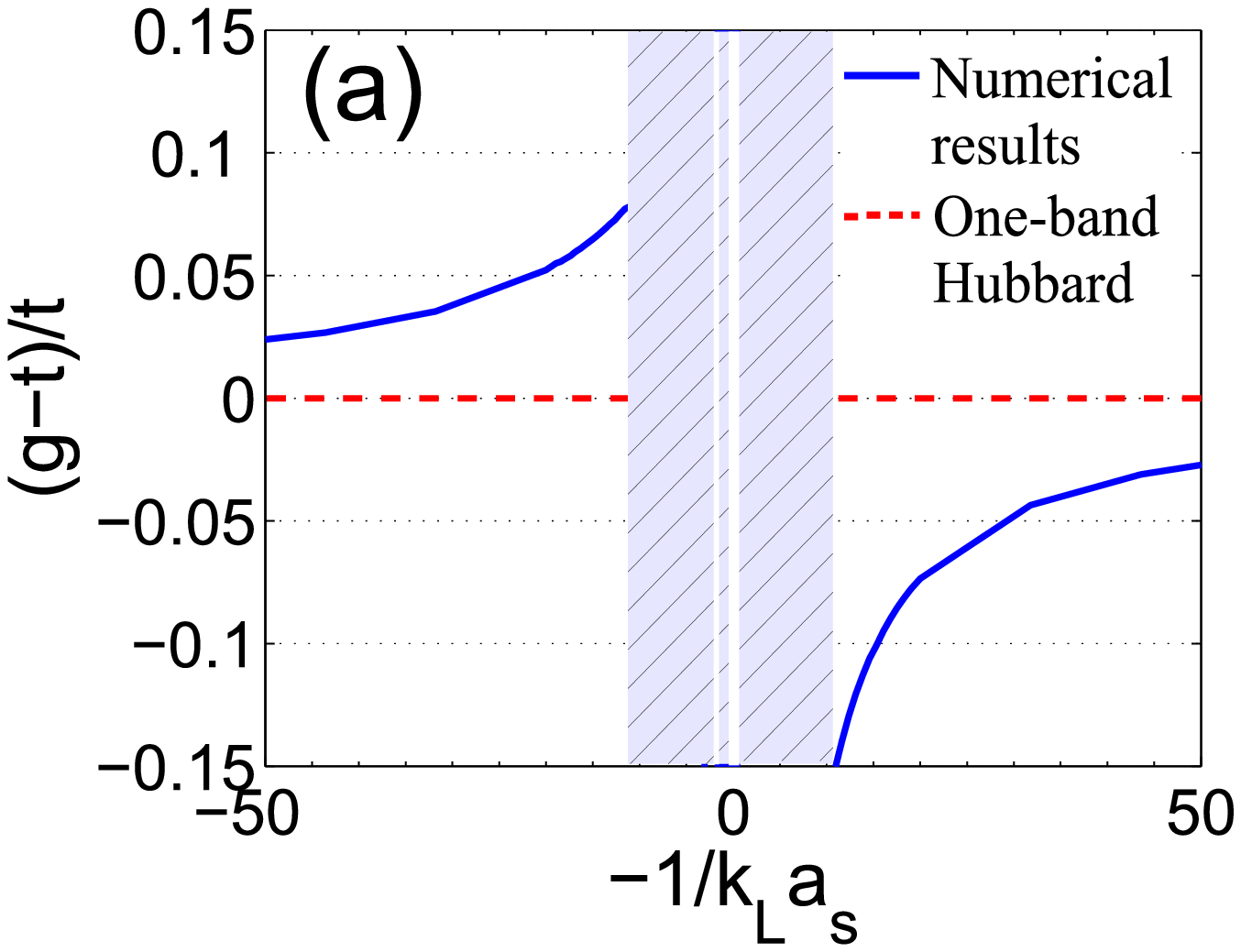}}
\subfigure  {\includegraphics[width=1\columnwidth]
        {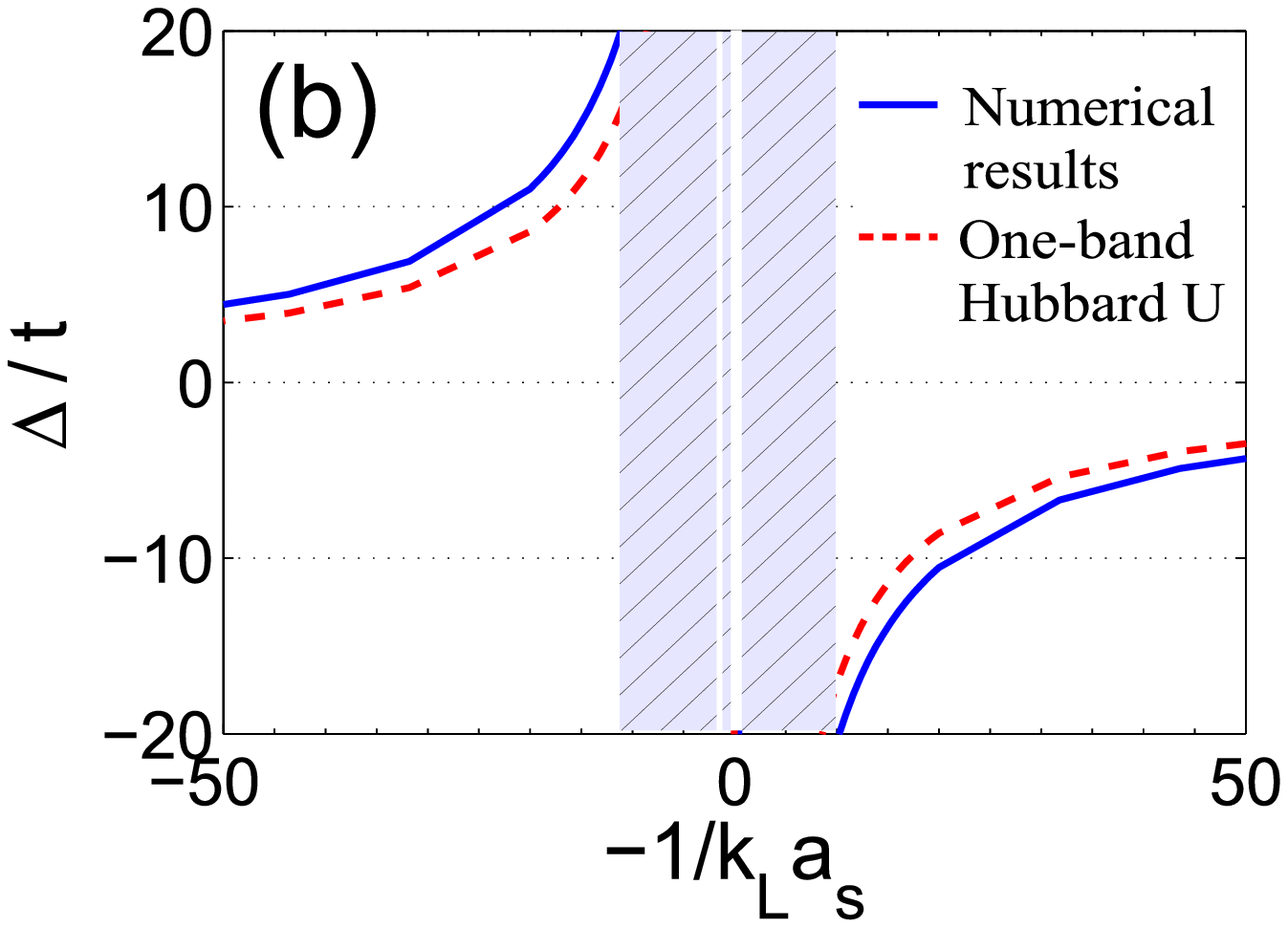}}
\caption{(Color online.) (a) Particle-assisted hopping rate and (b) on-site interaction energy vs inverse scattering
length for $V_{0}=8E_{R}$. Dimer hopping is negligible and not shown.} \label{fig:GHMparams}
\end{figure}
Above we have determined the structure of the relevant model Hamiltonians across unitarity. Now we consider the details
of the Hamiltonians and their parameters. Neglecting off-site interactions, the general Hubbard model takes the form
\cite{Duan08}
\begin{multline}
H_{GHM}=\sum_{i,\sigma }\left( n_{i\bar{\sigma}}\Delta -\mu \right) n_{i\sigma }  \label{eq:GHM1}
\\
-\sum_{i,j,\sigma}^{\prime }\biggl[t+\left( g-t\right) \left( n_{i\bar{ \sigma}}+n_{j\bar{\sigma}}\right)
\\
+\left( t_{da}+t-2g\right)
n_{i\bar{\sigma}}n_{j\bar{\sigma}}\biggr] a_{i\sigma }^{\dagger }a_{j\sigma }-t_{d}\sum_{i,j,\sigma }^{\prime
}a_{i\sigma }^{\dagger }a_{i\bar{\sigma}}^{\dagger }a_{j\bar{\sigma} }a_{j\sigma }
\end{multline}
where the prime on the sum means that only nearest neighbor terms are included, $\sigma $ denotes a fermion component
$\{\uparrow ,\downarrow \}$ and $\bar{\sigma}$ denotes the other component, $n_{i\sigma }=a_{i\sigma }^{\dagger
}a_{i\sigma }$, $\Delta $ is the on-site interaction energy, $\mu $ is the chemical potential, $g$ is the
particle-assisted tunneling rate, $ t_{da}$ is the rate for an atom to hop from a doubly occupied site to a singly
occupied site, and $t_{d}$ is the dimer tunneling rate. For just two atoms on two sites, the Hamiltonian can be written
as the matrix
\begin{equation}
H_{GHM}^{\left( 2,2\right) }=
\begin{pmatrix}
\Delta  & -g & -g & -t_{d} \\ -g & 0 & 0 & -g \\ -g & 0 & 0 & -g \\ -t_{d} & -g & -g & \Delta
\end{pmatrix}
-2\mu I  \label{eq:GHM2}
\end{equation}
in the basis $\{|\uparrow \downarrow ,0\rangle ,|\uparrow ,\downarrow \rangle ,|\downarrow ,\uparrow \rangle
,|0,\uparrow \downarrow \rangle \}$.

From the spectrum we can extract the model parameters as a function of the
experimental parameters by choosing them such that the lattice model correctly reproduces the four energy levels. The
resulting values depend on the scattering length and are shown in Fig.~\ref{fig:GHMparams} for $ V_{0}=8E_{R}$. The
model is not valid in the hatched regions, in accord with Fig.~\ref{fig:spectrum4}. The particle-assisted hopping rate,
$g$, generally differs from the single-particle hopping rate, $t$, by about $10\%$. (We use the value of $t$ obtained
from the one-atom spectrum.) Away from resonance, the on-site interaction energy, $\Delta $, is $25\%$ larger than the
standard Hubbard $U$ computed from the overlap of the lowest Wannier functions, but this is no doubt due in part to the
additional confinement compared to the infinite lattice case. We have not shown the dimer hopping rate away from
resonance, as it is negligible -- less than $1\%$ of the single-particle hopping rate in this case. Far from unitarity,
where the interaction is too weak to populate higher bands, the general Hubbard model reduces to the one-band Hubbard
model. In general, except for within the slivers of model validity around the anharmonicity induced resonances \cite{Kestner09}, Fig.~\ref{fig:GHMparams} shows that the general Hubbard model is qualitatively similar to the standard Hubbard model with some quantitative corrections as one begins to approach the strongly interacting region.

\begin{figure}[tbp]
\subfigure  {\includegraphics[width=.49\columnwidth]
        {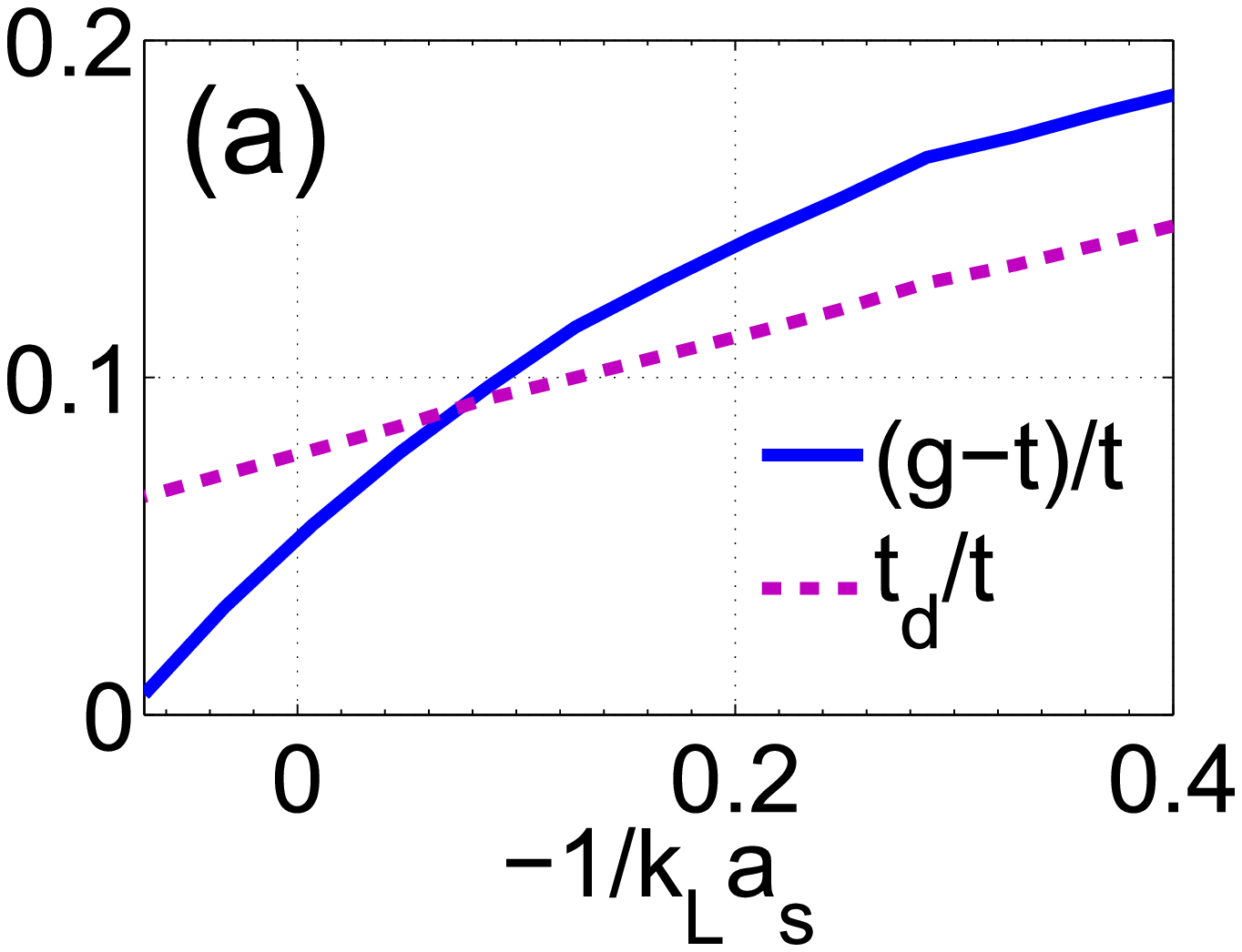}}
\subfigure  {\includegraphics[width=.49\columnwidth]
        {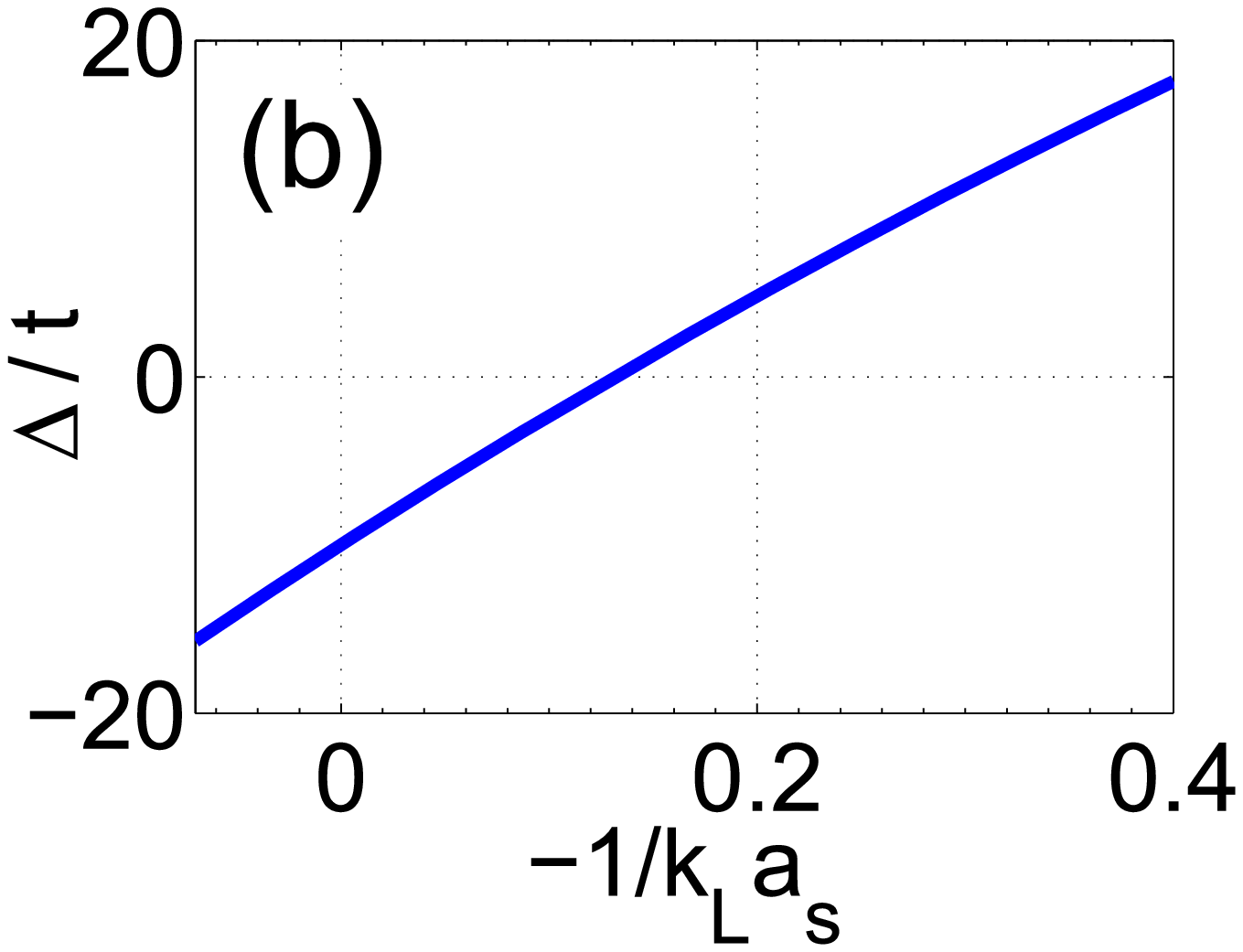}}
\caption{(Color online.) (a) Particle-assisted hopping rate, dimer hopping rate, and (b) on-site interaction energy vs
inverse scattering length in the narrow region around the first anharmonicity induced resonance where a general Hubbard
model is valid for $V_{0}=8E_{R}$.} \label{fig:GHMparamsres}
\end{figure}

Around the first anharmonicity induced resonance, though, the general Hubbard model is strikingly different than the
single-band Hubbard model, as shown in Fig.~\ref{fig:GHMparamsres}. Of course, the physical single-band approximation is not expected to hold in this strongly interacting region, but it is still interesting to contrast the two models. Most
notably, the effective on-site interaction does not become unbounded near the free space Feshbach resonance. Although
the relevant on-site dressed molecule state resembles the lowest dressed molecule state in that it is tightly bound
internally, it is only slightly detuned energetically from the atom pair singlet state due to its excited center of mass motion. Also, the dimer tunneling is no longer negligible, as might be expected for an excited dimer in a relatively
weak lattice. Only the particle-assisted hopping is qualitatively similar to that plotted away from unitarity in
Fig.~\ref{fig:GHMparams}.
\begin{figure}[tbp]
\subfigure  {\includegraphics[width=.49\columnwidth]
        {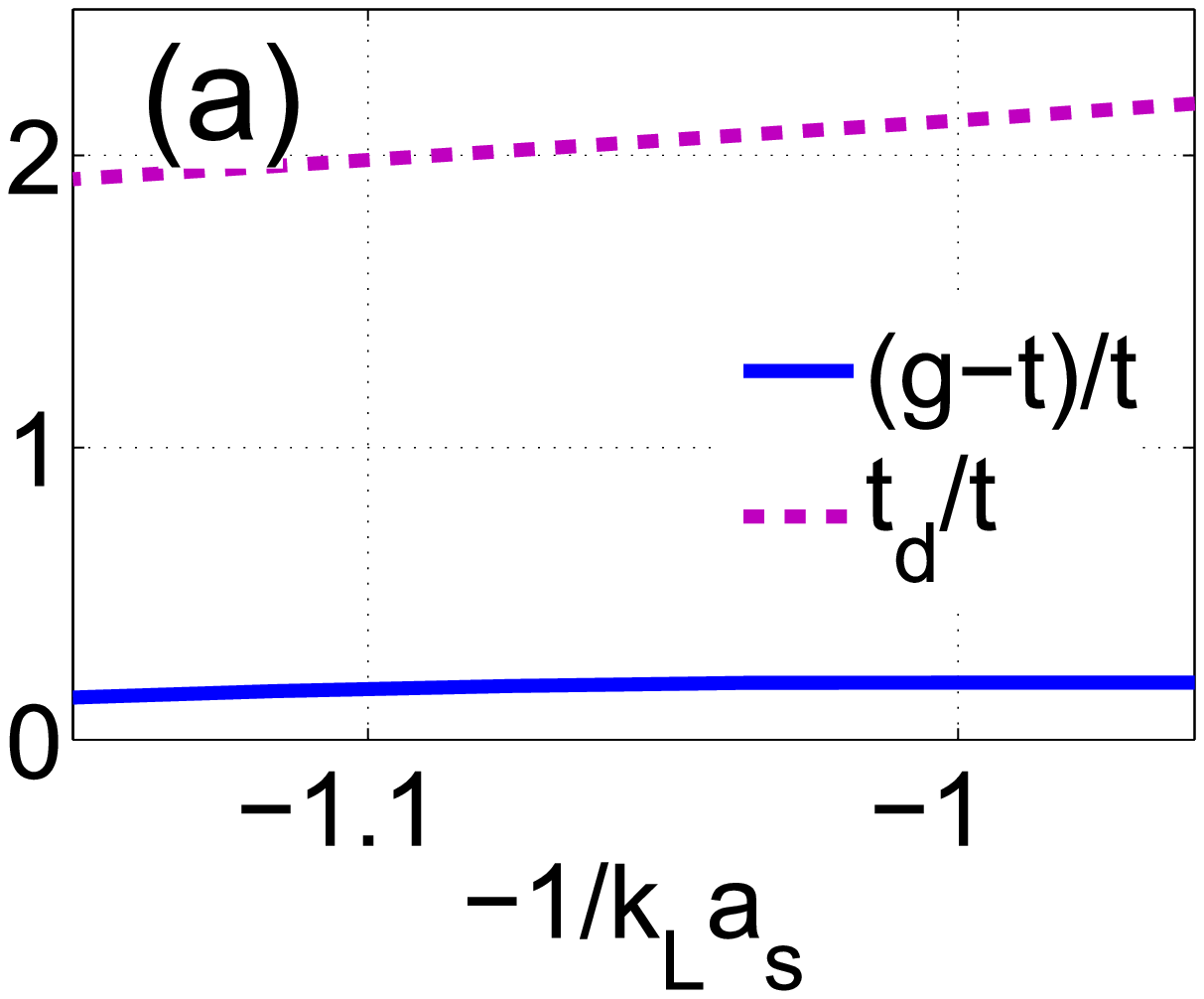}}
\subfigure  {\includegraphics[width=.49\columnwidth]
        {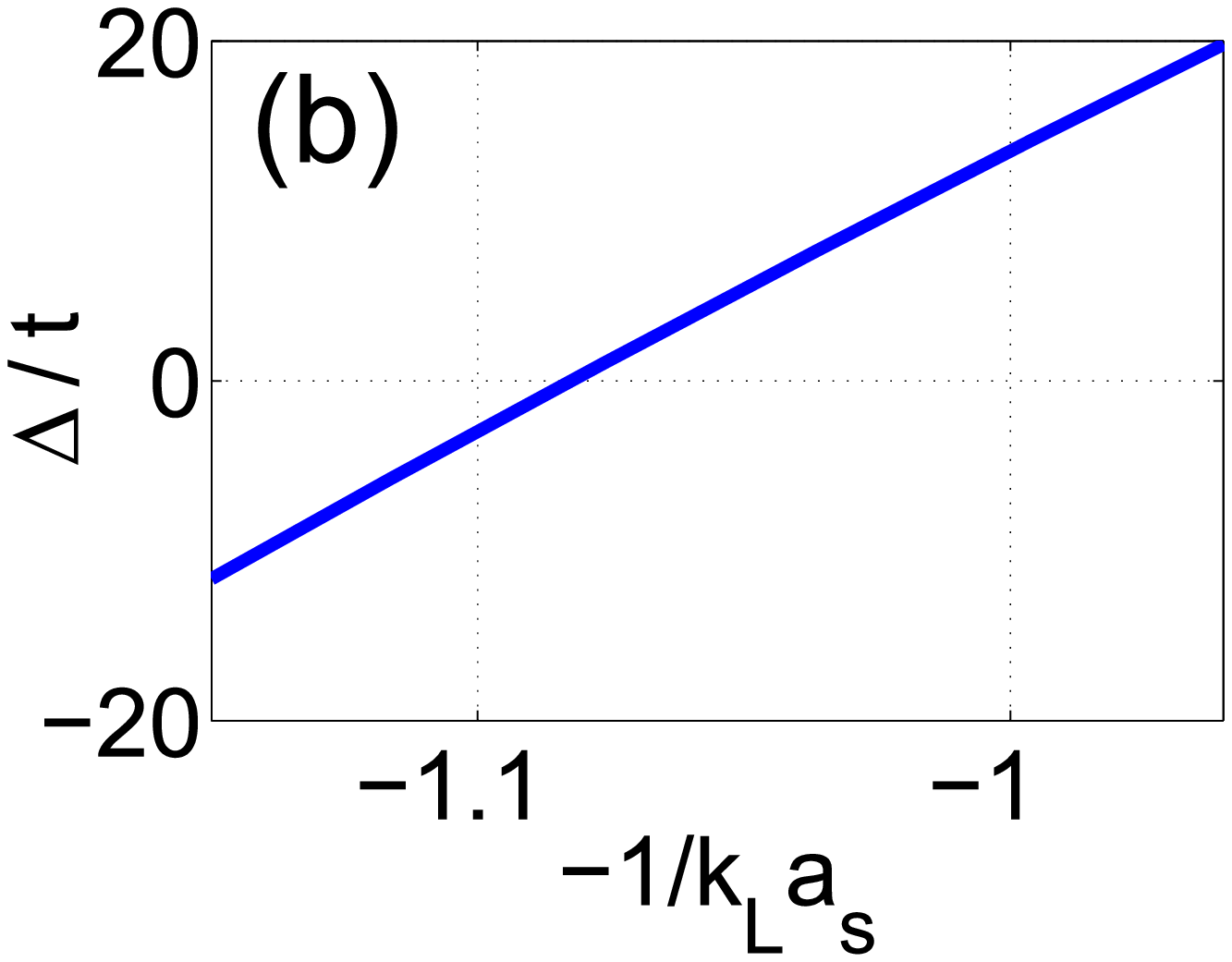}}
\caption{(Color online.) Same as Fig. \protect\ref{fig:GHMparamsres} but in the narrow region of model validity around
the second induced resonance.} \label{fig:GHMparamsres2}
\end{figure}

Similarly, for the small region around the crossing of the second excited molecule state and the lowest atomic state,
one can obtain the general Hubbard model parameters shown in Fig.~\ref{fig:GHMparamsres2}. The particle-assisted hopping and the on-site interaction energy are much the same as in Fig.~\ref{fig:GHMparamsres}. However, here the dimer
tunneling rate is much larger, twice as large as the atom tunneling rate, due to the excited nature of the relevant
molecule state.

\begin{figure}[tbp]
\includegraphics[width=1\columnwidth]{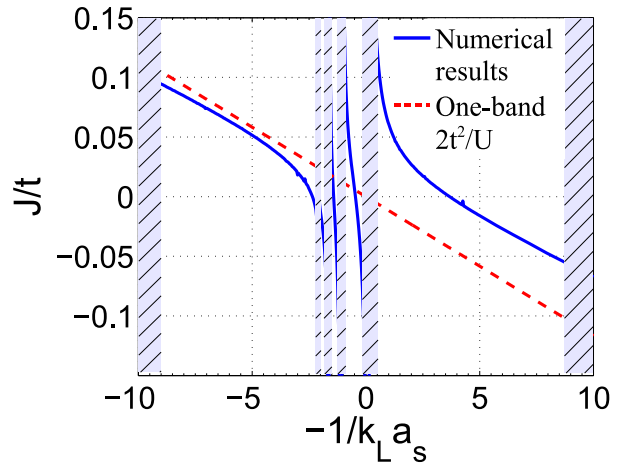}\newline
\caption{(Color online.) Superexchange energy vs. inverse scattering length for $V_{0}=8E_{R}$.} \label{fig:tJparams}
\end{figure}

The effective $t-J$ model has the familiar form
\begin{multline}
H_{t-J}=-\mu \sum_{i,\sigma }n_{i\sigma }  \label{eq:tJ1} \\ -\sum_{i,j}^{\prime }\left[ t\sum_{\sigma }P_{a}a_{i\sigma
}^{\dagger }a_{j\sigma }P_{a}-J\left( \mathbf{s_{i}}\cdot \mathbf{s_{j}} -n_{i}n_{j}/4\right) \right] ,
\end{multline}
where $P_{a}$ is a projector onto the subspace with at most one atom per site, $J$ is the superexchange energy,
$n_{i}=\sum_{\sigma }n_{i\sigma }$, and $\mathbf{s_{i}}=\sum_{\sigma ,\sigma ^{\prime }}a_{i\sigma }^{\dagger }
\mathbf{\sigma }_{\sigma ,\sigma ^{\prime }}a_{i\sigma ^{\prime }}/2$, with $ \mathbf{\sigma }$ denoting the Pauli
matrices. For just two atoms on two sites, the Hamiltonian can be written as the matrix
\begin{equation}
H_{t-J}^{\left( 2,2\right) }=-
\begin{pmatrix}
J & J \\ J & J
\end{pmatrix}
-2\mu I  \label{eq:tJ2}
\end{equation}
in the basis $\{|\uparrow ,\downarrow \rangle ,|\downarrow ,\uparrow \rangle \}$. We extract from the spectrum the
effective superexchange energy relevant for the regions where a $t-J$ model is valid. We have plotted this in
Fig.~\ref{fig:tJparams}, along with the result of $2t^{2}/U$ one would derive from the conventional one-band Hubbard
model. Again, this is just for reference, as there is no expectation that the physical single-band approximation is
valid in the strongly interacting regime. It is interesting to note that $J$ does not vanish at unitarity, but in fact
passes through zero on both sides of unitarity (in fact, more than once on the positive-$ a_{s}$ side). This implies
that one may find some transition to an exotic phase there, where the interaction is dominated by higher-order or
next-nearest neighbor processes. A similar zero-crossing has previously been found on the positive $a_{s}$ side of
resonance in a completely different calculation \cite{Mathy09}.

As the lattice depth is adjusted, the behavior of the parameters remains qualitatively unchanged and the maximum values
of the particle-assisted hopping rate and the superexchange energy remain more or less the same. The only exception is
that $t_d/t$ decreases quickly as the lattice becomes deeper, as expected. Where both models are valid, the $t-J$ model
can be derived from the general Hubbard model and $J = 2 g^2/\Delta $.

\section{Conclusions}

We have discussed the basic form of candidate effective single-band lattice Hamiltonians to describe the low-energy
physics of ultracold fermionic atoms in an optical lattice. General considerations of the relevant Hilbert space and
system symmetry \cite{Duan08} lead to two possibilities for energies near the non-interacting ground state: an effective single-band generalized Hubbard model and a $t-J$ model.

We have performed numerical calculations of the spectrum of two interacting fermions in a double-well potential to
determine under what conditions one of these lattice models is a good description of a physical system in a periodic
optical potential. By requiring the lattice models to reproduce the low-energy two-site two-atom physics in their
respective regions of validity, we have determined the on-site and nearest-neighbor parameters of the models. We find
that at unitarity there exists a valid effective single-band Hubbard model with counter-intuitive weak on-site
interaction. We also find that near unitarity there exists a valid $t-J$ model whose superexchange energy can be tuned
through zero on either the attractive or repulsive side of resonance. These models should prove useful starting points
for future theoretical or experimental considerations of strongly correlated many-body physics in an optical lattice. In particular, the ability to tune $J$ through zero suggests an interesting phase diagram in the vicinity of the crossing.

This work was supported by the AFOSR\ through the MURI quantum simulation program, the DARPA OLE program, the IARPA, and the ARO\ through a MURI\ program.

\end{document}